\documentclass[12pt,preprint]{aastex}

\def\hub{\ifmmode H_\circ\else H$_\circ$\fi}

\newcommand{\noprint}[1]{}

\usepackage{lscape,graphicx,afterpage}
\begin{document}

\title{Entropy of Intermediate-Mass Black Holes}

\author{Paul H. Frampton}

\affil{University of North Carolina at Chapel Hill}

\affil{Department of Physics and Astronomy, CB3255, Chapel Hill, NC 27599}

\email{frampton@physics.unc.edu}

\begin{abstract}
\noindent Observational searches for Intermediate-Mass Black Holes (IMBHs), 
defined to have masses 
between 30 and 300,000
solar masses, provide limits which 
allow up to ten percent of what is presently
identified as halo dark matter to be in the form of IMBHs. 
These concentrate entropy so efficiently that the
halo contribution can be bigger than 
the core supermassive black hole.
Formation of IMBHs is briefly discussed.  

\end{abstract}

\section{Introduction}

\noindent Surely the most spectacular confirmed prediction
of general relativity is the black hole.
Black holes are generally
characterized by mass, charge and angular
momentum. In the present paper I shall focus on mass 
because the electric charges of black holes in Nature
are likely to be extremely small while angular momentum will 
here be suppressed for no better reason than simplicity.

\noindent I shall first introduce acronyms for three mass ranges of
black hole in terms of the solar mass ($M_{\odot}$).
Supermassive black holes (SMBHs) 
have $M_{SMBH} > 3 \times 10^5 M_{\odot}$.
Intermediate-mass black holes (IMBHs) 
are defined by $3 \times 10^5 M_{\odot} > M_{IMBH}
> 30 M_{\odot}$. Because most
black holes are in the mass range I designate as 
Normal-mass black holes (NMBHs)
$30M_{\odot} > M_{NMBH}
> 3M_{\odot}$.

\noindent The compelling evidence for the existence
of both SMBHs and NMBHs is well known and need not be repeated here.
The status of IMBHs is reviewed in \citep{mc04}
with the conclusion
that their existence in Nature is not absolutely certain.
In the
present paper, I shall suggest 
on the basis of entropy arguments
\footnote{In an otherwise comprehensive discussion 
it is noteworthy 
that the word {\it entropy}
occurs only once in \citep{mc04}}  
that many IMBHs exist
in galactic halos and it is the job of
astronomers to detect them.

\section{Entropy Preamble}

\noindent Entropy has at least two interesting
properties. Firstly there is probably an upper limit
\footnote{The holographic principle has not, to my knowledge,
been rigorously proven.}
which arises from the holographic principle
\citep{ho94}.
Second, there
is the idea that it never decreases, the
second law of thermodynamics.

\noindent I shall employ throughout dimensionless entropy
which means that Boltzmann's formula
is divided by his eponymous constant. 
Entropy is a powerful concept
in theoretical physics generally and 
may have found a ubiquitous application 
in cosmology.

\noindent In discussing the dimensionless entropy
of the universe I encounter very large
numbers like a googol ($10^{100}$) and 
it will become clear how numbers below
one googol are negligible. In discussing
the number of microstates of the universe
any number below one googolplex ($10^{10^{100}}$)
is insignificant.

\section{Entropy of Everything Except Black Holes}

\noindent Black holes focus entropy density 
so efficiently that the entropy of everything else in the universe
is negligible compared to its black holes.

\noindent Since this is crucial in justifying  my
focus on black holes, let me begin by considering
briefly the entropy of other components which contribute
to the energy of the universe. These are  approximately
$72\%$ dark energy, $24\%$ dark matter and $4\%$ normal matter. 

\noindent I shall throughout make the assumption that the
dark energy possesses no entropy 
\footnote{If dark energy entropy were discovered, 
the discussion would be modified.}.

\noindent It is convenient to introduce the notation
where $10^x$ denotes within one order of magnitude.
$10^x$ means an integer greater than $10^{x-1}$ and less
than $10^{x+1}$.

\noindent Normal matter, other than relic photons and neutrinos,
has entropy $10^{80}$ nowhere near to one googol. The photons of
the Cosmic Microwave Background (CMB) have 
entropy $10^{88}$, orders of magnitude bigger, still negligible.
The relic neutrinos have entropy comparable in order
of magnitude to the CMB.

\noindent What about dark matter? Unlike dark energy it must possess statistical
properties like temperature and entropy and, other than black holes,
its dimensionless entropy must fall far short of one googol.

\noindent I conclude that black holes overwhelmingly dominate
cosmological entropy.

\begin{figure}
\plotone{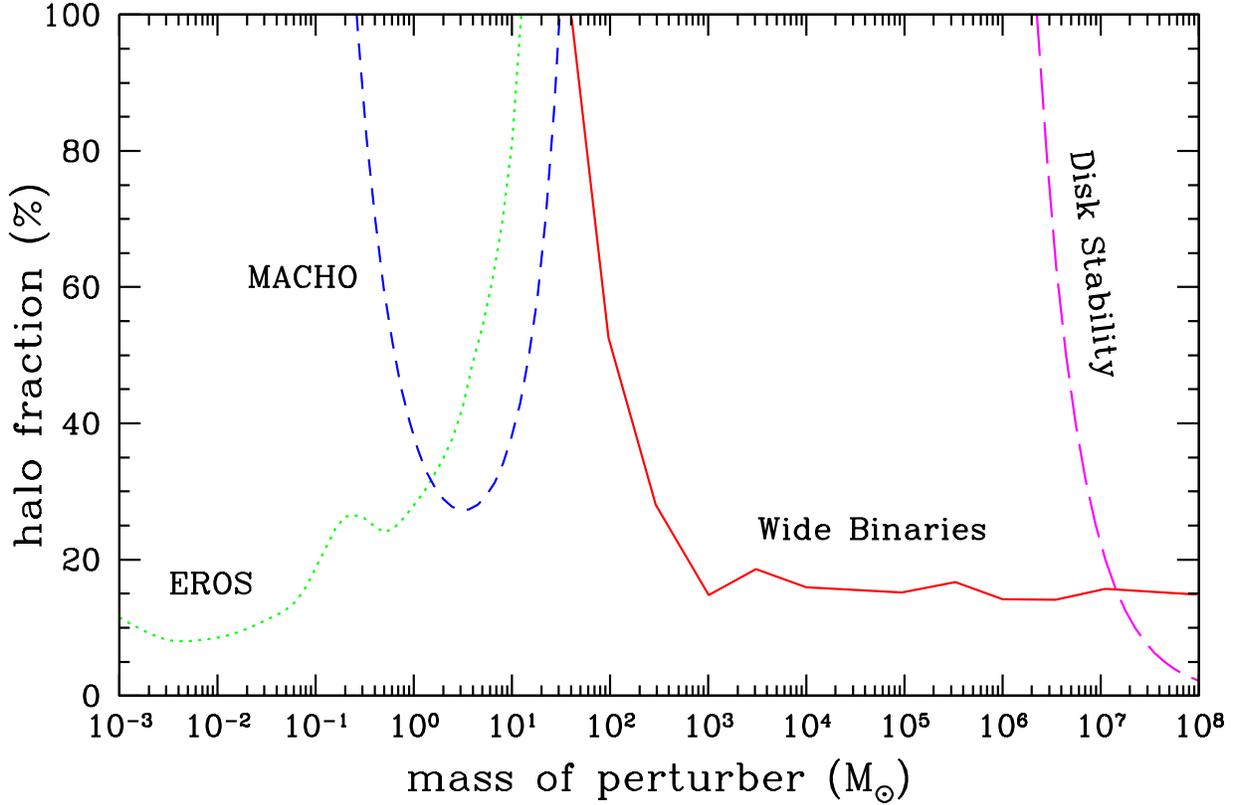}
\caption{The vertical axis is the total intermediate-mass black hole (IMBH)
mass as the percent of the halo dark matter; the
horizontal axis is the individula IMBH mass in terms of the solar
mass $M_{\odot}$.
Note that for the range of masses between $30M_{\odot}$ and 
$3 \times 10^5 M_{\odot}$ as much as ten percent of the
halo mass can be IMBHs, as can be seen by drawing a horizontal
line at 10\% and noting that it avoids the disallowed regions
for all $30M_{\odot} < M_{IMBH} < 3 \times 10^5 M_{\odot}$. Taken from \citep{Yoo04}}
\label{fig:frampton1}
\end{figure}

\section{Entropy of Black Holes}

\noindent From work \cite{pa69,be73,ha74}
(PBH) beginning
four decades ago 
it is well established that
the dimensionless entropy $S_{BH}$ of a black hole with mass $M_{BH} 
= \eta M_{\odot}$ is $10^{78}\eta^2$.

\noindent As a first (unrealistic) example consider the Sun
which is normal matter with entropy $10^{57}$. Now
take a black hole with the same mass: following the
PBH analysis it has entropy
$10^{78}$. Although the Sun is too light
ever to collapse to a black hole, this gedankenexperiment
illustrates the compression of entropy by 21 orders
of magnitude such that in any entropy inventory with
black holes everything else can be jettisoned with impunity.

\noindent Observed stellar NMBHs are at least three times
the mass of the Sun and have masses between $3M_{\odot}$
and $30M_{\odot}$. For the sake of simplicity I assume
every NMBH has $10M_{\odot}$ and therefore
entropy $10^{80}$. 

\noindent I shall assume the visible universe contains
$10^{11}$ galaxies each with halo mass $10^{12}M_{\odot}$.

\noindent As a working estimate of
the total entropy of NMBHs, I assume there
are $10^{10}$ of them per galaxy and hence
$10^{21}$ in the universe. Their total PBH entropy
is $10^{101} = 10$ googols.

\noindent Let me turn now to the SMBHs at galactic centers.
From the Southern Hemisphere in the direction of the
constellation Sagittarius, where the Milky Way appears
densest to the naked eye, there is good radio
astronomical evidence
for a SMBH known as Sag $A^{*}$ with mass 
of about four million solar masses. For simplicity I assume
there is one SMBH with mass $10^7M_{\odot}$
in every galaxy. Each has entropy $10^{92}$ and
their total cosmological entropy is therefore
$10^{103} = 10^3$ googols.

\noindent Finally I arrive at my principal topic,
the intermediate-mass black holes whose observational
evidence is inadequate,
As an example let me take all IMBHs to
have mass $10^5M_{\odot}$ and entropy $10^{88}$.
According to Fig. 1 as much as 10 percent
\footnote{I am grateful to Gianfranco Bertone
for bringing \cite{Yoo04} to my attention.}
of the $10^{12}M_{\odot}$
halo can be IMBHs so there can be $10^{6}$ per galaxy
and $10^{17}$ in the universe giving cosmological entropy
of $10^{105}=10^5$ googols and $99\%$ of total entropy.

\noindent The holographic principle 
allows a total entropy of the universe up to $10^{123} = 10^{23}$
googols achievable only if the universe were to
become one big black hole quite different from
the present situation.

\section{Cyclicity}

\noindent One reason that it is so urgent 
to seek and discover IMBHs is that their contribution
to cosmological entropy is germane to a better understanding
of the correct alternative to the big bang. 

\noindent In the cyclic model of (Baum \& Frampton 2007;
Baum \& Frampton 2008)
which solves
the Tolman conundrum, the total entropy provides a lower
limit on the number of universes spawned at turnaround
(Baum, Frampton \& Matsuzaki 2008)
and is relevant to the infinite past
(Frampton 2007; Frampton 2009). The cyclic 
model 
can solve purely classically the concern expressed in 
(Penrose 2005)
about fine-tuning to less than a part in a googolplex
at the start of an expansion era.

\section{Formation of IMBHs}

\noindent There are two formation mechanisms which suggest themselves:

\noindent Firstly, the IMBHs may be remnants of
Population III stars as discussed in \cite{mr01}.
If so,
the mass estimates for Population III
stars have been too conservative and they may range up to 
$3 \times 10^5 M_{\odot}$.

\noindent  Secondly, numerical simulations 
\cite{nfw96} of
dark matter halos could be tightened
to much higher spatial resolution and include
realistic inelasticity or equivalent dynamical friction.
In this way IMBH formation could be studied assiduously.
Note that the largest IMBH has a radius less than
$10^{-7}$ parsecs so the improvement
must be by orders of magnitude.

\noindent It is possible that IMBH formation
results sequentially from both mechanisms.

\section{Discussion}

\noindent The situation with respect to
cosmological entropy
is as if with respect to cosmological
energy there remained
ignorance of dark matter and dark energy.

\noindent I have put stock in 
the second law of thermodynamics to predict
that entropy is dominated by IMBHs. The
applicability of the second law
to the evolution of the universe ignores
dynamics yet I believe
the dynamics would need to be pathological 
if IMBHs do not prevail in the galactic halo.

\noindent From Fig. 1, Wide Binaries (Chaname \& Gould 2004) offer 
an opportunity to tighten the constraint on, 
or to discover, IMBHs.

\noindent The MACHO searches \cite{al97,al00}  
found many definite examples of microlensing
all with longevities less than 2y. 

\begin{center}

{\bf Table I}

\begin{tabular}{||c||c||}
\hline\hline
$ \log_{10} \eta$  &  $t_0$ (years) \\
 \hline\hline
3 &  6 \\
     \hline
4 & 20 \\
      \hline
5 &  60 \\
 \hline\hline
\end{tabular}
         
\noindent {\bf Intermediate Mass Black Holes (IMBHs) and 
Microlensing Longevity.}

\noindent {\it Here $\eta$ is defined by the mass of the IMBH being 
$\eta M_{\odot}$ where $M_{\odot}$ is the solar mass 
and $t_0$ is the longevity of a hypothetical
microlensing event produced by the putative IMBH.}

\end{center}

\noindent From Table I 
we see, however, that higher microlensing longevities 
are pertinent to IMBHs. For the example of 
$M_{IMBH} = 10^5 M_{\odot}$,
it is 60y.

\noindent The Disk Stability upper limits on $M_{IMBH}$ seem 
robust due to the useful and important 
work in \cite{lo85,xo94}.

\noindent On the theoretical side the observed
metallicity of intergalactic dust requires
that Population III stars exist and little is
established about their masses. I suggest that
there may have once existed far higher-mass zero-metallicity
stars than discussed in \cite{mr01}. 

\noindent For numerical simulations \cite{nfw96}
of the dark matter
halo to become useful for
the study of IMBH formation they must dramatically
improve on spatial resolution. 

\noindent In conclusion, I believe that the
IMBH entropy make up of the universe can be understood
in a just few years. My question is for 
astronomers. What is the percentage contribution
by intermediate-mass black holes to the
total cosmological entropy?

\begin{center}

{\bf Acknowledgement}

\end{center}

\noindent
This work was supported by U.S. Department of Energy 
grant number DE-FG02-06ER41418.

\end{document}